\documentclass[12pt]{article}
\usepackage{amsfonts,epsfig,latexsym,amscd}
\usepackage{graphics}
\usepackage{epsfig}
\usepackage{graphicx}
\newcommand{\be}{\begin{equation}}
\newcommand{\ee}{\end{equation}}
\newcommand{\bea}{\begin{eqnarray}}
\newcommand{\eea}{\end{eqnarray}}

\newcommand{\fr}{\frac}

\newcommand{\ra}{\rightarrow}
\newcommand{\al}{\alpha}
\newcommand{\bt}{\beta}
\newcommand{\pr}{\partial}
\newcommand{\hs}{\hspace{5mm}}

\newcommand{\dg}{\dagger}

\newcommand{\acc}{\\[3mm]}

\newcommand{\bp}{\mbox{\boldmath$\phi$}}
\newcommand{\bn}{\mbox{\boldmath$n$}}

\begin{document}
\begin{center}
{\bf Discrete  Skyrmions in 2+1 and 3+1 Dimensions}\acc
T{\small HEODORA} I{\small OANNIDOU}$^\dag$\footnote{{\it Permanent Adress:} School of Mathematics, Physics and Computational Sciences, 
   Faculty of Engineering,
Aristotle University of Thessaloniki, Thessaloniki 54124, Greece}  and 
 P{\small ANOS} K{\small EVREKIDIS}$^\ddag$\\
$^\dag${\it TAT, Eberhard Karls Universitat T\"ubingen, 72076 T\"ubingen,
 Germany}\\  
$^\ddag${\it Department of Mathematics and Statistics, University of
 Massachusetts, Amherst, MA $01003-4515$, USA}\\
{\it Emails: ti3@auth.gr\\\hspace{40mm} kevrekid@math.umass.edu} 
\end{center}

This paper describes a lattice version of the Skyrme model in $2+1$ and 
$3+1$ dimensions.  The discrete model is derived from a consistent 
discretization of the radial continuum problem.
Subsequently, the existence and stability of the skyrmion solutions 
existing on the lattice are investigated. One consequence of the proposed 
models is that the corresponding discrete skyrmions have a high degree of 
stability, similar to their continuum counterparts.

\section{Introduction}

Many attempts have been made in order to obtain the discrete analogue of 
given  continuum systems that admit solitary wave 
solutions, so that  their characteristics in the continuum 
be preserved in the lattice. 
One such characteristic for the topological systems is often
the existence of
the so-called Bogomolnyi bound \cite{bog}; while for the integrable
ones it may be the existence of the Lax pair \cite{AS}.
In the continuum, the stability of the topological solitons is often 
related to the existence of the  energy bound while, the stability of the 
solitons in the integrable systems  is related  to the presence of  an 
infinite number of conservation laws.
However, in the lattice, only in a few  cases  the Bogomolnyi bound has  
been preserved 
\cite{L,SW,I} while it is even more difficult to define the lattice
Lax pair (especially in higher dimensions) of the corresponding 
integrable system \cite{AS}. Moreover, the stability of the solitons in 
question  is  not guaranteed. Often, under discretization, the 
topological properties are essentially lost (as may be expected 
since topology is related to continuity).
Lattice versions of these nonlinear wave bearing 
systems have been much studied (for purposes of numerical 
simulations \cite{ahs1} or regularization of the quantum field theory, or
because they are of fundamental physical interest in their own 
right \cite{fl1}).

For a given continuum model, there are  many different discrete analogues 
which reduce to it in the continuum limit. The object of this paper it to
 present a lattice version of the Skyrme model in $2+1$ and $3+1$ dimensions.
The Skyrme model is a popular model of the dynamics of pions and nucleons,
 incorporating the former as its fundamental pseudo-Goldstone field and the 
latter as topological solitons. Its continuum version has been widely studied
 using numerical and analytical methods (for more details, see for example,
 Ref. \cite{SM}).
Its lattice formulation is of some importance since the model is
 non-renormalizable in perturbation theory and thus, existing treatments of
 the model are semiclassical (quantizing only the collective degrees of 
freedom of the soliton). Full quantization of the theory requires a cutoff
 which can be attained by its lattice version.

The most interesting feature of the Skyrme model is the stability conferred
 on the soliton by the topology \cite{Za}. An open question is whether this 
stability is preserved on the lattice. In this paper, lattice skyrmions 
obtained within an appropriate discretization of the Skyrme model 
in $2+1$ and $3+1$ dimensions are typically found to be stable in our 
parametric investigations. This suggests that these discretizations bear
some important features of their continuum counterparts, while being
easier to handle from a numerical point of view.

Our presentation is arranged as follows. The next section reviews the
 baby Skyrme model and then reparametrizes the fields to impose radial
 symmetry. Only then is the model discretized, as shown in section $2.1$;
 while in section $2.2$,
numerical results on the 
the existence and stability of a single (radially symmetric) baby skyrmion on 
the lattice are given. The case of soliton configurations in the Skyrme model
 in $3+1$ dimensions was dealt with in the same way in section 3. Finally,
 our conclusions are presented in section 4.

\section{The Baby Skyrme Model}

Let us begin with a brief review of the  Skyrme  model in $2+1$-dimensions (so-called   baby Skyrme model).
The Lagrangian density of the  model is of the form
\be 
{\cal L}=\fr{1}{2}\pr_\al \bp\, \pr^\al \bp-
\fr{\kappa}{4}(\pr_\al\bp\times \pr_\bt \bp)(\pr^\al\bp\times \pr^\bt \bp)
-\mu^2\left(1-(\bn\cdot \bp)^2\right).\label{bs}
\ee
The field $\bp$  is a map from $M^3\ra S^2$, where $M^3$ 
is the $3$-dimensional Minkowski space with  metric $\eta^{\al\bt}=\rm{diag}(1,-1,-1)$
 and the target space $S^2$ is the $2$-sphere of unit radius embedded in 
Euclidean $3$-space.
Therefore, the field $\bp$ is a scalar $3$-vector with norm one, i.e. 
$|\bp|^2=1$. The constants  $\kappa, \mu$ are free parameters which have
 the dimension of length and energy, respectively.
The first term in (\ref{bs}) is the familiar $O(3)$ sigma model, the
 second  term is the $2$-dimensional analogue of the Skyrme term and 
the last term is  the potential. 

The presence of the potential in (\ref{bs}) ensures the existence of 
stable skyrmion solutions.
There are two other options in literature for its form: 
 i) the holomorphic model where the potential term is $(1+\bn\cdot \bp)^4$ but
  stable skyrmions cannot be obtained and  (ii) the old baby Skyrme model
 where the  potential term is  $(1-\bn\cdot \bp)$ and stable  non-radially
 symmetric skyrmions  exist. 
The Lagrangian (\ref{bs}) corresponds to the new baby Skyrme model 
\cite{Tom} which possess radially symmetric skyrmions.
In \cite{IKZ}, approximate analytic skyrmion solutions of the new baby Skyrme model were 
obtained by exploring its topological properties.

Finiteness of the energy requires the potential term to vanish at infinity, implying that $\lim_{r \ra \infty}\bp(t,x,y)=\bn$ (where 
$r=\sqrt{x^2+y^2}$). For simplicity
we choose $\bn$ to be the vacuum state, that is  $\bn=(0,0,1)$.
Thus  a topological number exists  since  the field $\bp$, due to the boundary conditions, can be considered
 as a map  from $S^2\ra S^2$. 

The topological charge is the homotopy
 invariant of the field 
\be
{\rm deg}[\bp]=\fr{1}{4\pi}\int\bp \cdot (\pr_x\bp\times \pr_y\bp)\,dx\,dy.
\ee 
 and thus, conserved.

\subsection{Discrete Baby Skyrmions}

In order to obtain the discrete analogue of the baby Skyrme model 
 we restrict our considerations to fields which are invariant under 
simultaneous rotations and reflections in space and target space. Thus, we
 assume that the field $\bp$ is of the hedgehog form
\be \phi_i=k_i \sin g(r,t)k_i,\hs \phi_3=\cos g(r,t)  \ee
where $k_i$ for $i=1,2$ is a unit vector given in terms of the azimuthal 
angle $\theta$  and the topological charge $N=\deg[\bp]$
as $k_i=(\cos N\theta,\sin N\theta)$; and $g(r,t)$ is the real profile
 function
which satisfies certain boundary conditions. Then,  the respective energy  
functionals related to Lagrangian (\ref{bs}) are
\bea
E_{\rm kin}&=&\pi\int r\dot{g}^2\left(1+\fr{\kappa^2N^2}{r^2}\sin^2 g\right)
 dr\label{ek}\\
E_{\rm pot}&=&\pi\int \left(rg_r^2+\fr{\kappa^2N^2}{r}g_r^2\sin^2 g
+\fr{N^2}
{r}\sin^2 g+\mu^2 r \sin^2 g\right) dr.\label{ep}
\eea
The  boundary conditions for the skyrmions are: $g(0,t)=\pi$ and $g(r,t)=0$ as 
$r\ra \infty$. Note that the first and  third term in (\ref{ep})  
corresponds to the static $O(3)$ sigma model energy; 
the second term corresponds to the $2$-dimensional static Skyrme  energy  and the last one is the potential.

Hereafter, $r$ becomes a discrete variable with lattice spacing $h$. So,
the real-valued field $g(r,t)$ depends on the continuum variable $t$
 and the
discrete variable $r=nh$ where  $n\in Z^+$. Then, $g_+=g((n+1)h,t)$
 denotes forward shift and thus, the forward difference is given by
 $\Delta g=(g_+-g)/h$. There are many possibilities for discretizing the
 above energy functionals. However, based on the approach introduced 
 in \cite{I} for  the discretization of the $O(3)$ sigma model, we assume that
\bea
g_r&=&\fr{2f(h)}{h}\sin\left(\fr{g_+-g}{2}\right)\nonumber\\
\sin g&=&\fr{1}{f(h)}\sin\left(\fr{g_++g}{2}\right).
\label{lat}
\eea
The parameter $f(h)$ is an arbitrary function of the lattice spacing 
subject to the
 constraint $f(h)\ra 1$ as $h\ra 0$. 

The origin must be treated in a special way since the functionals are not 
defined at $n=0$. One possibility is to assume  (following \cite{I}) 
that at the origin we have 
\bea
\left(rg_r^2\right)\Big{|}_{r=0}&=&\left(\fr{N^2}{r}\sin^2 g\right)
\Big{|}_{r=0}\nonumber\\
&=&\fr{2N}{h}\cos^2\left(\fr{g(h,t)}{2}\right).\label{ori}
\eea

Then, the kinetic and potential energy of the discrete baby Skyrme model
 are defined by the following expressions
\bea
E_{\rm kin}&=&\pi\sum_{n=1}^\infty nh^2\dot{g}^2
\left\{1+\fr{k^2N^2}{n^2h^2f^2}
\sin^2\left(\fr{g_++g}{2}\right)\right\}\nonumber\acc
E_{\rm pot}&=&4\pi N\cos^2\left(\fr{g(h,t)}{2}\right)\left[1+
\fr{2\kappa^2f^2}{h^2}\cos^2 \left(\fr{g(h,t)}{2}\right)\right]\nonumber\\
&+&\pi\sum_{n=1}^\infty\left\{4nf^2\sin^2\left(\fr{g_+-g}{2}\right)+
\fr{\kappa^2N^2}{nh^2}\sin^2\left(\fr{g_+-g}{2}\right)\sin^2
\left(\fr{g_+ +g}{2}\right)\right. \nonumber\\
&&\left.\hs\hs\,\,\, +\fr{N^2}{nf^2}\sin^2\left(\fr{g_+ +g}{2}\right)
+\fr{\mu^2nh^2}
{f^2}\sin^2\left(\fr{g_+ +g}{2}\right)
\right\}.\label{len}
\eea
Note that, for discretizing the second term of the energy at the origin,
  a combination of formula (\ref{lat}) and (\ref{ori}) has been used. 
Actually, when this term is absent  no stable lattice baby skyrmions can
 be obtained.
Recall, that  this term is the discrete analog of the Skyrme term which 
 stabilizes the solution
and apparently its presence  is vital (even at the origin).

For $\kappa=\mu=0$ the model (\ref{len}) becomes the 
discrete version of the
 $O(3)$ sigma model introduced in \cite{I}.  

The lattice equations of motion obtained from the
 variation of the Lagrangian $L=E_{\rm kin}-E_{\rm pot}$ given by   (\ref{len}) are
\bea
&&\ddot{g}\left[1+\fr{\kappa^2N^2}{h^2f^2}\sin^2\left(\fr{g_++g}{2}\right)
\right]+\fr{\kappa^2N^2}{2h^2f^2}\sin(g_++g)\left(\fr{\dot{g}^2}{2}+
\dot{g}\dot{g_+}\right)\nonumber\\
&&=\fr{N}{h^2}\sin g\left(1+\fr{4\kappa^2f^2}{h^2}\cos^2\fr{g}{2}\right)
+\sin (g_+-g)\left[\fr{g^2}{h^2}+\fr{\kappa^2 N^2}{4h^4}\sin^2\left(
\fr{g_++g}{2}\right)\right]\nonumber\\
&&\hs-\sin(g_++g)\left[\fr{N^2}{4h^2f^2}+\fr{\mu^2}{4f^2}+\fr{\kappa^2N^2}{4h^4}
\sin^2\left(\fr{g_+-g}{2}\right)\right],\,\,\, n=1\nonumber
\acc
&&n\ddot{g}\left[1+\fr{\kappa^2N^2}{nh^2f^2}\sin^2\left(\fr{g_++g}{2}\right)
\right]+\fr{\kappa^2N^2}{2h^2f^2}
\left[\fr{\sin(g_++g)}{n}\left(\fr{\dot{g}^2}{2}+
\dot{g_+}\dot{g}\right)-\fr{\sin(g+g_-)}{n-1}\fr{\dot{g}_-^2}{2}\right]
\nonumber\\
&&=-\sin(g-g_-)\left[\fr{g^2}{h^2}(n-1)+\fr{\kappa^2N^2}{4h^4}\fr{1}{(n-1)}
\sin^2\left(\fr{g+g_-)}{2}\right)\right]\nonumber\\
&&\hs+\sin(g_+-g)\left[\fr{g^2}{h^2}\,n+\fr{\kappa^2N^2}{4h^4}\fr{1}{n}\sin^2
\left(\fr{g_+ +g}{2}\right)\right]\nonumber\\
&&\hs-\sin(g_++g)\left[\fr{N^2}{4f^2h^2}\fr{1}{n}+\fr{\mu^2}{4f^2}n+
\fr{\kappa^2 N^2}
{4h^4}\fr{1}{n}\sin^2\left(\fr{g_+-g}{2}\right)\right]\nonumber\\
&&\hs-\sin(g+g_-)\left[\fr{N^2}{4f^2h^2}\fr{1}{(n-1)}+\fr{\mu^2}{4f^2}(n-1)+
\fr{\kappa^2N^2}{4h^4}\fr{1}{(n-1)}\sin^2\left(\fr{g-g_-}{2}\right)\right],
 \,\,\, n>1.\nonumber\\
\label{212}
\eea
Next, our task  is to study whether the aforementioned lattice equations admit skyrmion solutions and if this is the case  investigate whether the 
topological  properties of the solutions are maintained in the 
presence of the lattice.

\subsection{Numerical Simulations}

\begin{figure}[tbp]
\begin{center}
\epsfxsize=9.0cm
\epsffile{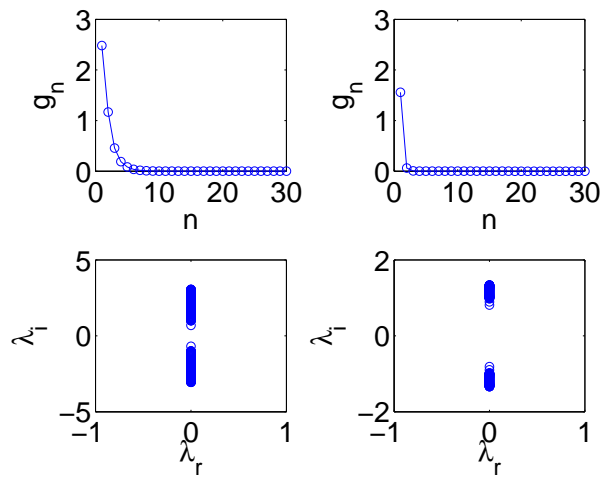}
\caption{
Typical profiles and stability results for the discrete baby
skyrmions.  The profile of the solution
is shown in the top panels for $h=0.65$ (left) and $h=1.5$ (right).
The corresponding spectral plane $(\lambda_r,\lambda_i)$ of
the eigenvalues $\lambda=\lambda_r+i \lambda_i$ of the linearization 
around   the solutions is shown in the bottom panels.}
\label{fig1}
\end{center}
\end{figure}

Our numerical procedure for obtaining the baby skyrmions 
 is the following: we use a fixed point iteration to identify
the static solutions of equations (\ref{212}).
An initial guess (for the fixed-point Newton iteration) in the
form of an inverse trigonometric function (an $\arccos$ in the
radial direction)  is used which  subsequently, after a few iteration steps,
converges to an exact stationary solution. Examples of such 
solutions are shown in Figure \ref{fig1} for $h=0.65$ (left panels) and
$h=1.5$ (right panels).
The results are obtained for the choices: 
$N=\kappa=\mu=f=1$ (unless noted otherwise), 
although variations of the parameters do not 
significantly affect our conclusions presented below. 

\begin{figure}[tbp]
\begin{center}
\epsfxsize=7.0cm
\epsffile{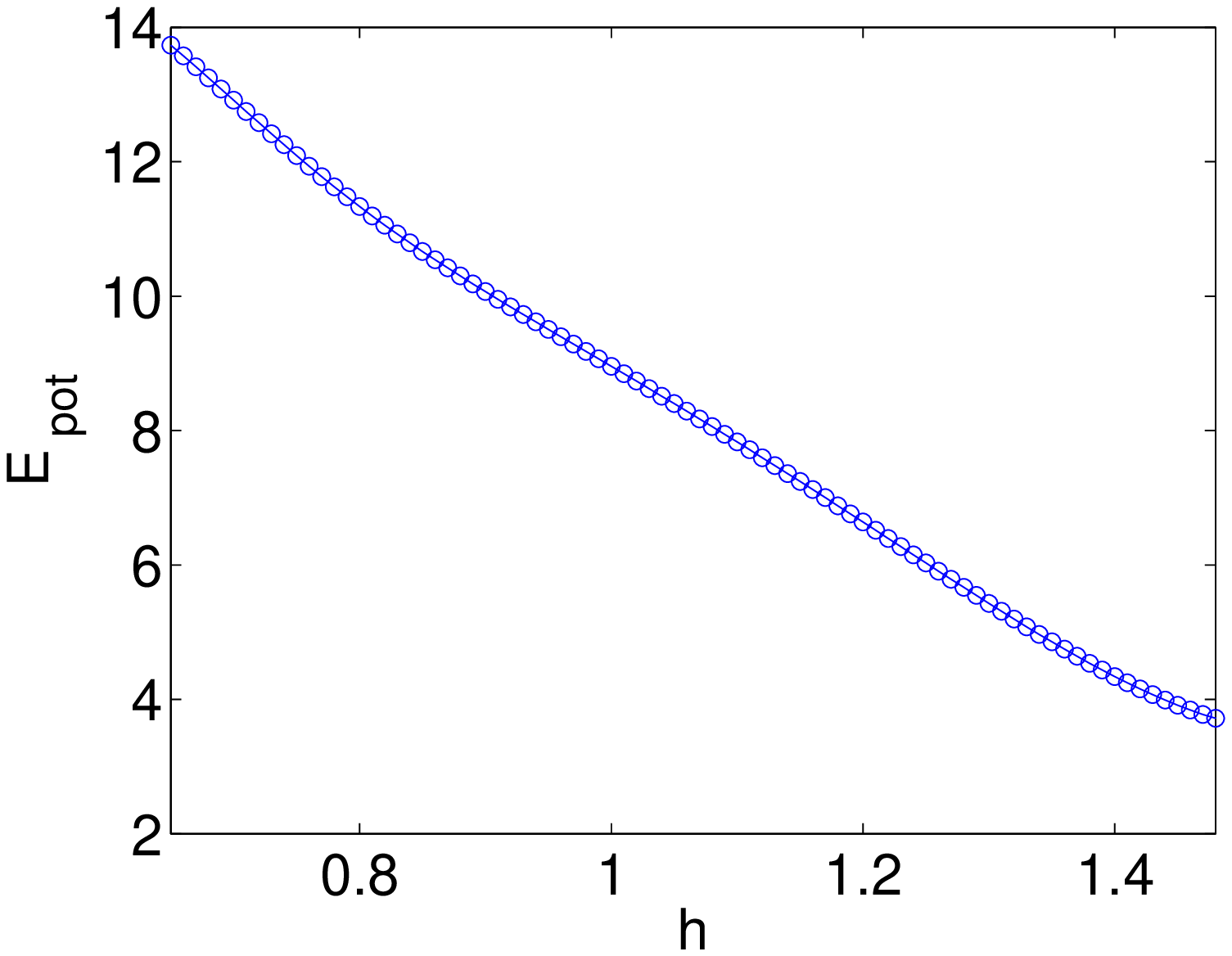}
\epsfxsize=7.0cm
\epsffile{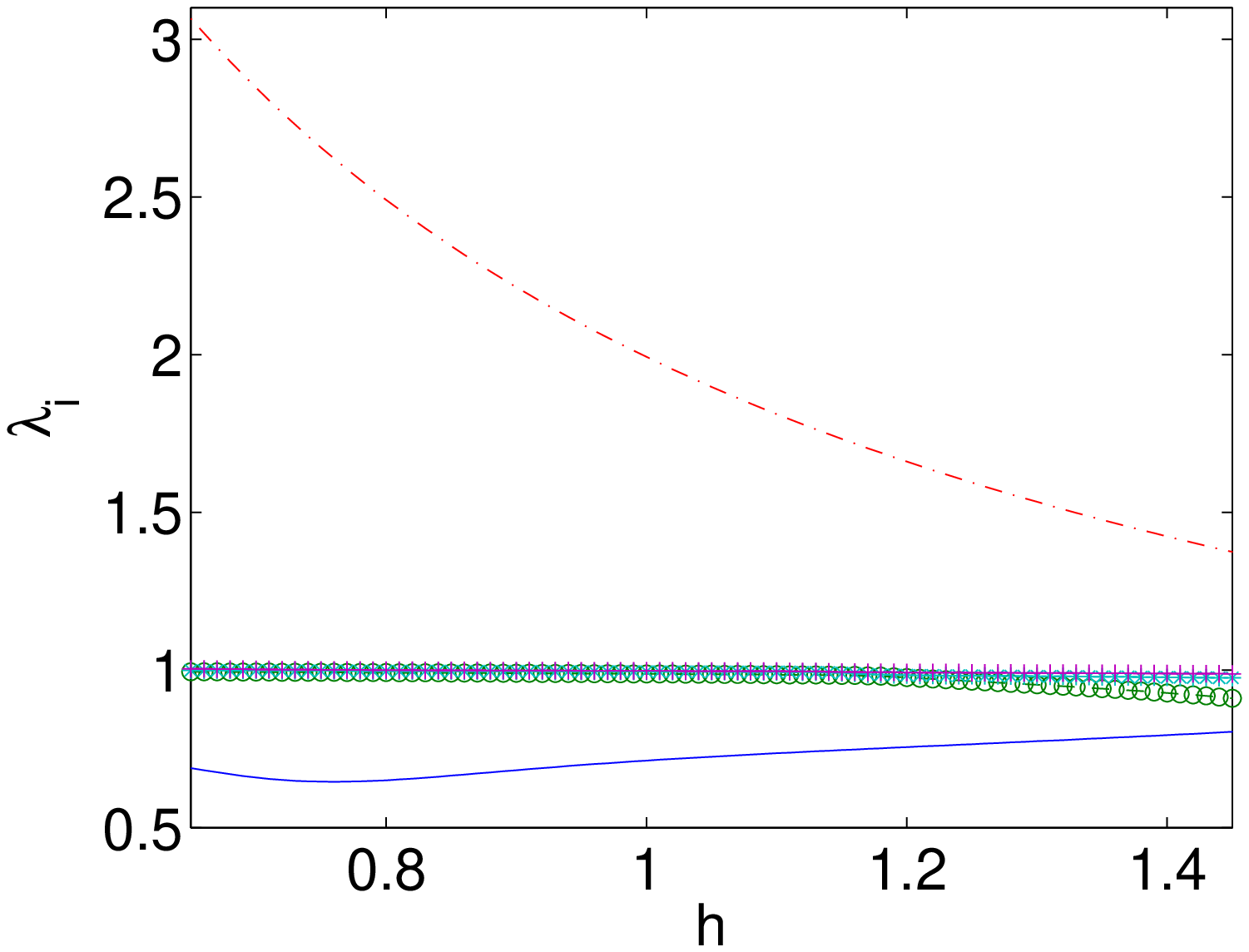}
\caption{The left panel of the figure shows the potential energy of our
exact static solutions given as a function of the lattice spacing
$h$. In the right panel the principal (minimal, as well as maximal)
imaginary parts of the 
eigenvalues of the linearization
around the solutions are shown as a function of $h$.}
\label{fig2}
\end{center}
\end{figure}

As expected, increase of $h$ renders the solution more ``coarse''
(i.e., with a fewer sites participating in the ``spine'' of the skyrmionic
structure). However, the stability of the wave is not crucially affected
by the discretization, since the baby skyrmion remains stable throughout
the range of parameters used in our numerical investigations. 
The stability is inferred from the eigenvalues of the relevant Jacobean 
of linearization around the solution. In particular, a
 linearization ansatz of the form:
\begin{eqnarray}
g_n=g_n^{st} + \epsilon \exp(\lambda t) w_n 
\label{stab}
\end{eqnarray}
leads to an eigenvalue problem (to O$(\epsilon)$, where $\epsilon$
is a formal small parameter) for the eigenvalue-eigenvector pair
$(\lambda,w_n)$; $g_n^{st}$ represents the stationary solution 
obtained
in the aforementioned Newton step. A solution is deemed to be stable
if none of the eigenvalues $\lambda=\lambda_r + i \lambda_i$ 
of the linearization problem is
found to have a strictly positive real part $\lambda_r$.
A particularly peculiar feature of the  
 model is that the second term on the left hand side of equations 
 (\ref{212}) does not contribute to the linearization
(its lowest order contribution is O$(\epsilon^2)$). This suggests that
even if the solution is found to be linearly stable, the full dynamics
of  the system (\ref{212}) should be considered, as it can, in 
principle, lead to nonlinear instabilities that cannot be detected
at the linearization level. The results of the branch of solutions and
their stability, obtained as a function of the lattice spacing $h$ are
 summarized
in Figure \ref{fig2}. The potential energy of the solutions,  evaluated
based on  (\ref{len}), reveals that the discretizations contributes
towards decreasing the energy of the stationary solutions.
The eigenvalues vary as a function of $h$, and interesting 
features such as bifurcations of internal modes (see e.g. \cite{cret,pgk}
and the references therein) arise e.g. for $h>1.15$ from the bottom 
edge of the continuous spectrum. Additionally, a point spectrum 
(isolated) eigenvalue
exists, which is also clearly discernible in the plots. Nevertheless, 
these do not appear to significantly affect the stability of the
obtained discrete baby skyrmion structures.

\begin{figure}[tbp]
\begin{center}
\epsfxsize=7.0cm
\epsffile{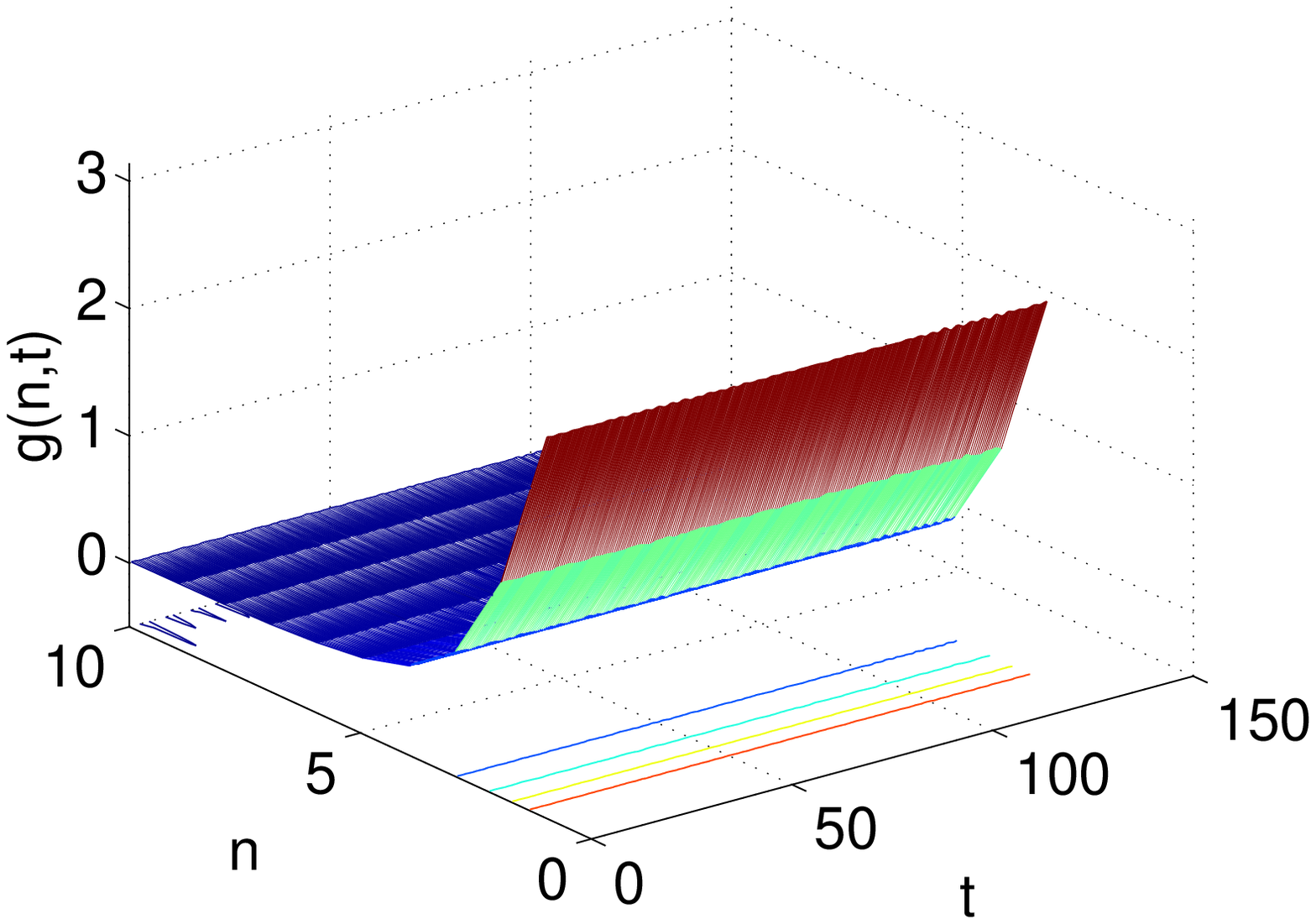}
\epsfxsize=7.0cm
\epsffile{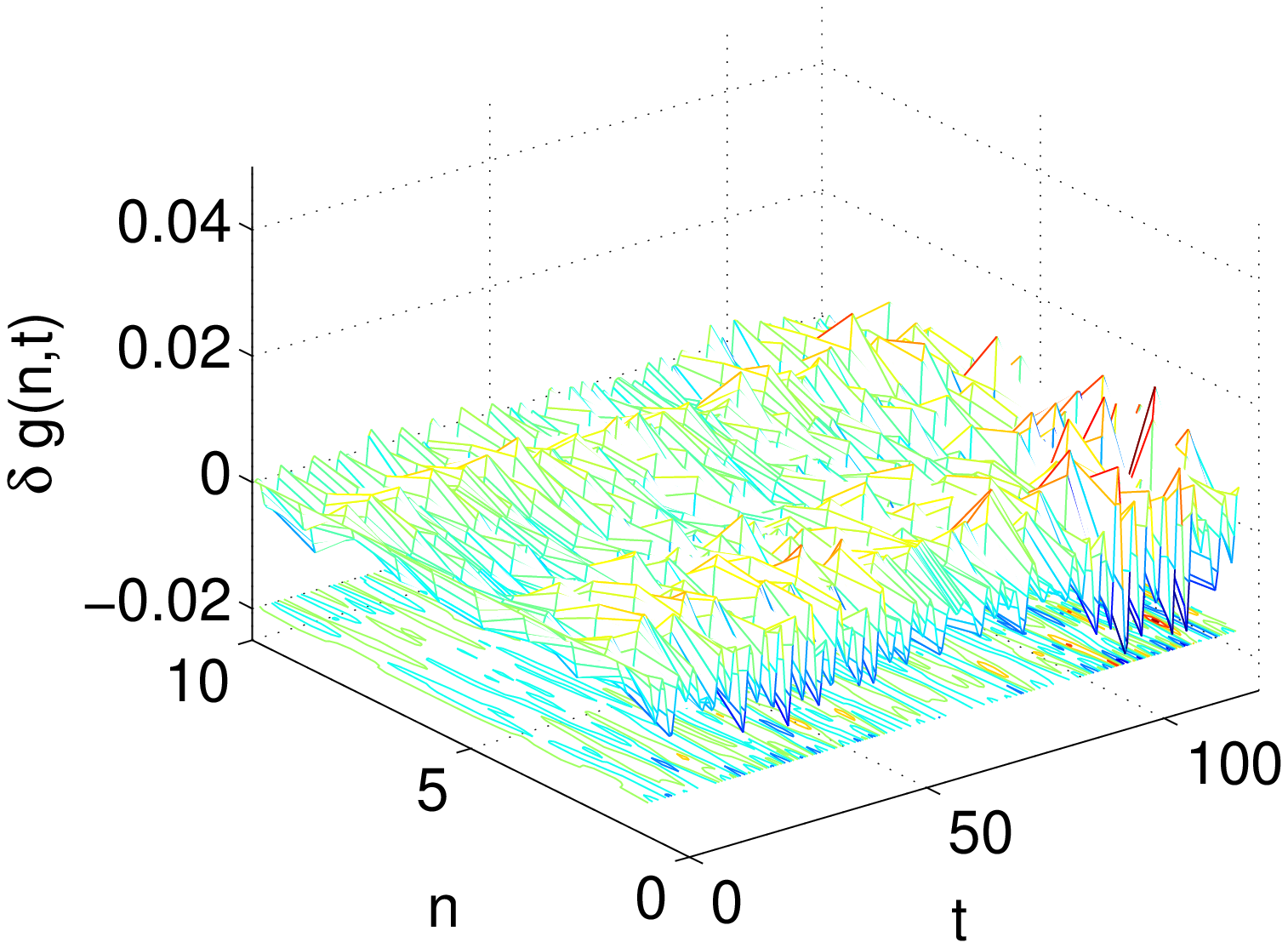}
\caption{Space-time evolution of the baby skyrmion for $h=0.65$
under an initial
(random, uniformly distributed) perturbation of amplitude $5 \times 10^{-3}$.
Note that despite the significant strength of the perturbation,
the solution profile remains essentially intact (left panel), while
the difference $\delta g(n,t)=g(n,t)-g(n,0)$ remains bounded (right panel)
attesting to the robust stability of the discrete 
baby skyrmion.}
\label{fig3}
\end{center}
\end{figure}

Finally, we examine the dynamical evolution of the baby skyrmions
to confirm the stability of the solutions obtained herein. We initialize
the full dynamics of  (\ref{212}) with an exact baby 
skyrmion
(for $h=0.65$ in Figure \ref{fig3}) perturbed by a random (uniformly 
distributed) field of amplitude $5 \times 10^{-3}$. 
In Figure \ref{fig3} the space-time evolution of the waveform and its
persistence in time (left panel) is presented, and it is obvious that despite
 the strength of
the perturbation (which is, however, substantial) remains bounded during
 its
time evolution in the right panel of the figure. This indicates that
the baby skyrmions are robustly stable dynamical structures of
the corresponding discrete equations.

Note that in \cite{Ward}, discrete $2$-dimensional topological skyrmions have been constructed for a novel lattice version of the baby Skyrme model (i.e., the $2$-dimensional topological Heisenberg model)     but their stability was far weaker than the continuum ones since a fairly small perturbation caused 
their decay.

\section{The  Skyrme Model in $3+1$ Dimensions}

The Lagrangian of the $SU(N)$ Skyrme model in $(3+1)$ dimensions can be
 written  in 
terms of the currents $R_\mu=\pr_\mu UU^{-1}$  as
\be
12\pi^2 {\cal L}=-\fr{1}{2}{\rm tr}\left(R_\mu R^\mu\right)
-\fr{1}{16}{\rm tr}\left([R_\mu,R_\nu] [R^\mu , R^\nu ]\right)
\ee
where we have used scaled units of energy and length, and a 
$(+,-,-,-)$
signature for the space-time metric.
The asymptotic value of the $SU(N)$ Skyrme field
$U(x,t)$ has to tend  to a constant matrix at spatial infinity, so that 
finite-energy configurations can exist.

The baryon density, whose spatial integral gives the integer-valued baryon 
number, is given by
\be
24\pi^2{\cal B}=-\varepsilon_{ijk}{\rm tr}\left(R_iR_jR_k\right).
\ee
From the mathematical point of view these field configurations represent
$3$-dimensional topological solitons.

The model is not integrable and thus, with few exceptions, explicit solutions
are hard to obtain. 
One way to overcome this problem is by expressing the Skyrme field in
 terms of
harmonic maps of $S^2$ to $CP^{N-1}$.
In particular, the idea of the rational map ansatz, introduced in \cite{HMS}
 is
 to separate the radial and angular dependence of the Skyrme fields.
  Its $SU(N)$  generalization introduced in \cite{IPZ}, expresses
 the Skyrme field  in terms of
 a profile function $g(r)$ and a $N\times N$  Hermitian projector $P$ that 
depends only on the angular variable as
\be 
U(r,\theta,\phi)=e^{2ig(r)(P-1/N)}.
\label{U}
\ee
The matrix $P$ is a harmonic map from $S^2$ into $CP^{N-1}$.
 Hence it is convenient
to map the sphere onto the complex plane via a stereographic projection in
 terms of the complex coordinate  $z=e^{i\phi}\tan(\theta/2)$ and its complex
conjugate.  In fact $P=(V\otimes V^\dg)/|V|^2$ is given in 
terms of a $N$-component complex vector dependent on $z$ and $\bar{z}$.
For (\ref{U}) to be well-defined at the origin, the radial profile function
has to satisfy $g(0)=\pi$ while as $r \rightarrow \infty$, it is
required that the Skyrme field 
has $\lim_{r \rightarrow \infty} g(r)=0$.

In \cite{HMS} it was shown that  ansatz (\ref{U}) describes field 
configurations for the $SU(2)$ model that are close to being solutions
of the model. In particular, although the corresponding energies are slightly
higher that the energies of the exact solutions (obtained numerically) the
symmetries of the baryon and energy densities are the same.
Also  in \cite{IPZ}, it was shown that when harmonic maps from $S^2$ to
$CP^{N-1}$  of the form (\ref{U}) are considered,  low-energy configurations of the $SU(N)$ Skyrme model can be
derived. These configurations 
are more symmetrical than the $SU(2)$ ones but have higher energies. 
However, exact solutions with spherically symmetry may also be  obtained from 
 (\ref{U}). 
They correspond to solutions of the  $SU(2)$ and $SU(3)$ Skyrme model with 
baryon number  $B=1$ and $B=0$ (topologically trivial solution), respectively.

Using (\ref{U}) the energy of the Skyrme model simplifies to
\bea
E_{\rm kin}&=&\fr{1}{3\pi}\int \dot{g}^2 (A_Nr^2+2{\cal N}\sin^2 g) dr\\
E_{\rm pot}&=&\fr{1}{3\pi}\int\left(A_Nr^2g_r^2+2{\cal N}(g_r^2+1)\sin^2g+
{\cal I}
\fr{\sin^4g}{r^2}\right) dr
\label{Sp}
\eea
where 
\bea
A_N&=&\fr{2}{N}(N-1)\\
{\cal N}&=&\fr{i}{2\pi}\int dz d\bar{z}\rm{tr}\left(|P_z|^2\right)\\
 {\cal I}&=&\fr{i}{4\pi}\int dz d\bar{z}(1+|z|^2)^2\rm{tr}
\left([P_z,P_{\bar{z}}]^2\right).
\eea
Note that the integrals ${\cal N}$ and 
${\cal I}$ are independent of $r$.
In particular, ${\cal N}$ corresponds to the energy of the $2$-dimensional 
$CP^{N-1}$ sigma model and is equal to the degree of the
highest-order polynomial in $z$ among the components of $V=R(z)$ 
(when being holomorphic) after all their common factors have been cancelled
 out.

Finally, the baryon number for this ansatz is
\be 
B=\fr{i}{2\pi}\int dz d\bar{z}\rm{tr}\left(P[P_{\bar{z}},P_z]\right)
\label{bar}
\ee
which is the topological charge of the $2$-dimensional $CP^{N-1}$ 
sigma model.

 In general, for $V$ holomorphic, it was proved in \cite{IPZ} that 
$B={\cal N}=n$ and ${\cal I}=n^2$   where $n={\rm deg}(V)$.
However,  for the special $SU(3)$  non-topological solution the vector $V$ is a function of $z$ and $\bar{z}$ (i.e. 
non-holomorphic) .  In this case, the baryon density is identically zero (${\cal B}=0$)  and the solution describes a bound state of two skyrmions and two antiskyrmions and as such is unstable as it corresponds to a saddle point of the energy. Let us emphasize that the field configuration is a genuine (non-trivial) solution of the $SU(3)$ Skyrme model; while  the corresponding  parameters are $A_N=\fr{4}{3}$ and  ${\cal N}={\cal I}=4$.

The Bogomolnyi-type argument for the Skyrme model gives the following lower energy bound
\bea
E_{\rm pot}\!\!\!&=&\!\!\!\fr{1}{3\pi}\int\left\{\left(\sqrt{A_N}rg_r+
\sqrt{{\cal I}}\,\,
\fr{\sin^2 g}{r}
\right)^2+
2{\cal N}\left(g_r+1\right)^2\sin^2 g\right.\nonumber\\
&&\hs\hs\hs\left. -2\left(\sqrt{A_N{\cal I}}+2{\cal N}\right)\sin g
 \pr_r(\cos g)\right\}dr\nonumber
\\
&\geq&\fr{1}{3}\left(2{\cal N}+\sqrt{A_N{\cal I}}\right),\label{smbog}
\eea
due to  the boundary conditions of the profile function.

Baby skyrmions are topological solitons of the field theory which resembles 
the  Skyrme model. Thus, in the following sections, we apply the  techniques developed for discretizing  the baby Skyrme model in the $3$-dimensional Skyrme model. 

\subsection{Discrete Skyrmions}

In what follows we present a lattice version of the Skyrme model by using the lower bound of the energy.
Thus, following \cite{SW} we start  with the same 
function $\sin g\pr_r(\cos g)$ as appears in (\ref{smbog}) and reconstruct
 the inequality
\be
\sqrt{A_N{\cal I}}\sin g\Delta\left(\cos g\right)=-D_ng_n\label{sx}\ee
where $D_n\ra \sqrt{A_N}rg_r$ and $F_n\ra \fr{\sqrt{\cal I}}{r} \sin^2 g$ in the
 continuum limit $h\ra 0$. The formula $\Delta(\cos g)=-2/h\sin
\left(\fr{g_+-g}{2}\right)\sin\left(\fr{g_++g}{2}\right)$ suggests the 
choices
\bea
D_n&=&\sqrt{A_N}(nh)\,\fr{2}{h}\,\sin\left(\fr{g_+-g}{2}\right)\nonumber\\
F_n&=&\sqrt{\cal I}\,\fr{1}{nh}\,\sin g\sin\left(\fr{g_++g}{2}\right).
\label{dfdis}
\eea
Also, the origin must be treated in a special way since (\ref{dfdis}) are 
undefined at $n=0$. One possibility is to arrange it so $D_0+F_0=0$ implying
 that
\be 
D_0=-F_0=\fr{(A_N{\cal I})^{1/4}}{\sqrt{h}}\sqrt{g(h,t)-\pi}\sin g(h,t)
\label{mid}
\ee
which follows for the discretization of the term: 
$\sqrt{A_N{\cal I}}\left[\Delta(g\sin^2 g)-g\Delta(\sin^2g)\right]$.
Note that, a direct discretization of (\ref{sx}) when 
 $n=0$ and assuming that $D_0=-F_0$ is not possible since the terms are  identically equal to zero due to the boundary condition $g(nh,t)|_{n=0}=\pi$.
 Thus, one possibility to overcome the problem  is to discretize its counterpart term
 $\pr_r(g\sin^2 g)-g\pr_r(\sin^2 g)$. This way the non-trivial ansatz (\ref{mid}) is obtained.

Then, the potential energy of the lattice  Skyrme model is given by
\bea
E_{\rm pot}&=&8\pi \sqrt{A_N{\cal I}}(g(h,t)-\pi)\sin^2 g(h,t)\nonumber\\
&+&4\pi h\sum_{n=1}^\infty\left\{4A_Nn^2\sin^2\left(\fr{g_+-g}{2}\right)+
\fr{{\cal I}}{n^2h^2}\sin^2g\sin^2\left(\fr{g_++g}{2}\right)\nonumber\right.
\\
&&\hs\hs\hs\,\,\left.+2{\cal N}\left[\fr{4}{h^2}\sin^2\left(\fr{g_+-g}{2}
\right)+1\right]\sin g\sin\left(\fr{g_++g}{2}\right)\right\}.\label{pesk}
\eea
Finally, the discrete version of the kinetic energy is
\be 
E_{\rm kin}=4\pi h \sum_{n}\left[A_N \,n^2h^2+2{\cal N} \sin g
\sin\left(\fr{g_++g}{2}\right)\right] \dot{g}^2.\label{kesk}
\ee

The corresponding Euler-Lagrange equations obtained from 
the Lagrangian $L=E_{\rm kin}-E_{\rm pot}$ 
where the potential and kinetic energy are,
respectively, given by   (\ref{pesk}) and (\ref{kesk}), read
\bea
&&\ddot{g}\left[A_Nh^2+2{\cal N}\sin g\sin\left(\fr{g_++g}{2}\right)\right]
+{\cal N} \dot{g}_+\dot{g}\sin g\cos \left(\fr{g_++g}{2}\right)
\nonumber\\
&&+{\cal N}\dot{g}^2\left[\cos g\sin\left(\fr{g_++g}{2}\right) +\fr{\sin g}{2}
\cos \left(\fr{g_++g}{2}\right)\right]\nonumber\\
&=&-\fr{\sqrt{A_N{\cal I}}}{h}\left[\sin^2 g+(g-\pi)\sin 2g\right]+ 
A_N \sin(g_+-g)\nonumber\\
&&-{\cal N}\left[\fr{4}{h^2}\sin^2\left(\fr{g_+-g}{2}\right)+1 \right]\left[\cos g
\sin\left(\fr{g_++g}{2}\right)+\fr{\sin g}{2} \cos\left(\fr{g_++g}{2}\right)\right]
\nonumber\\
&&+\fr{2{\cal N}}{h^2}\sin(g_+-g)\sin g\sin\left(\fr{g_++g}{2}\right)\nonumber\\
&&-\fr{\cal I}{2h^2}\left[\sin 2g \sin^2 \left(\fr{g_++g}{2}\right)+
\fr{\sin(g_++g)}{2}\sin^2 g\right], \hs n=1  \acc
&&\ddot{g}\left[A_Nn^2h^2+2{\cal N}\sin g\sin\left(\fr{g_++g}{2}\right)\right]
+{\cal N} \dot{g}_+\dot{g}\sin g\cos \left(\fr{g_++g}{2}\right)
\nonumber\\
&&+{\cal N}\dot{g}^2\left[\cos g\sin\left(\fr{g_++g}{2}\right) +\fr{\sin g}{2}
\cos \left(\fr{g_++g}{2}\right)-\fr{\sin g_-}{2} \cos \left(\fr{g+g_-}{2}\right)\right]\nonumber\\
&=&A_Nn^2\sin(g_+-g)-A_N(n-1)^2\sin(g-g_-)\nonumber\\
&&-{\cal N}\left[\fr{4}{h^2}\sin^2\left(\fr{g_+-g}{2}\right)+1 \right]\left[\cos g\sin\left(\fr{g_++g}{2}\right)+\fr{\sin g}{2} \cos\left(\fr{g_++g}{2}\right)\right]\nonumber\\
&&-\fr{\cal N}{2}\left[\fr{4}{h^2}\sin^2\left(\fr{g-g_-}{2}\right)+1 \right]\sin g_-\cos\left(\fr{g+g_-}{2}\right)\nonumber\\
&&-\fr{2{\cal N}}{h^2}\left[\sin(g-g_-)\sin g_- \sin\left(\fr{g+g_-}{2}\right)
-\sin(g_+-g)\sin g\sin\left(\fr{g_++g}{2}\right)\right]\nonumber\\
&&-\fr{\cal I}{2h^2}\left[\fr{\sin 2g}{n^2} \sin^2 \left(\fr{g_++g}{2}\right)+\fr{\sin(g+g_-)}{2(n-1)^2}\sin^2g_-+
\fr{\sin(g_++g)}{2n^2}\sin^2 g\right], \hs n>1. \nonumber \\
\label{stuff3d} 
\eea
In the next section, we will show that stable discrete skyrmion solutions of the aforementioned equations can be obtained numerically. In particular,  the analogues of the  $SU(2)$ $B=1$ and $SU(3)$ $B=0$ skyrmion configurations are constructed when the corresponding parameters are  $A_N=1$,
 ${\cal N}={\cal I}=1$ and $A_N=4/3$,  ${\cal N}={\cal I}=4$, respectively.

\subsection{Numerical Solutions}

\begin{figure}[tbp]
\centering
\hskip .2cm
\put(0,160){a)}
 \epsfxsize=11.0cm
\epsffile{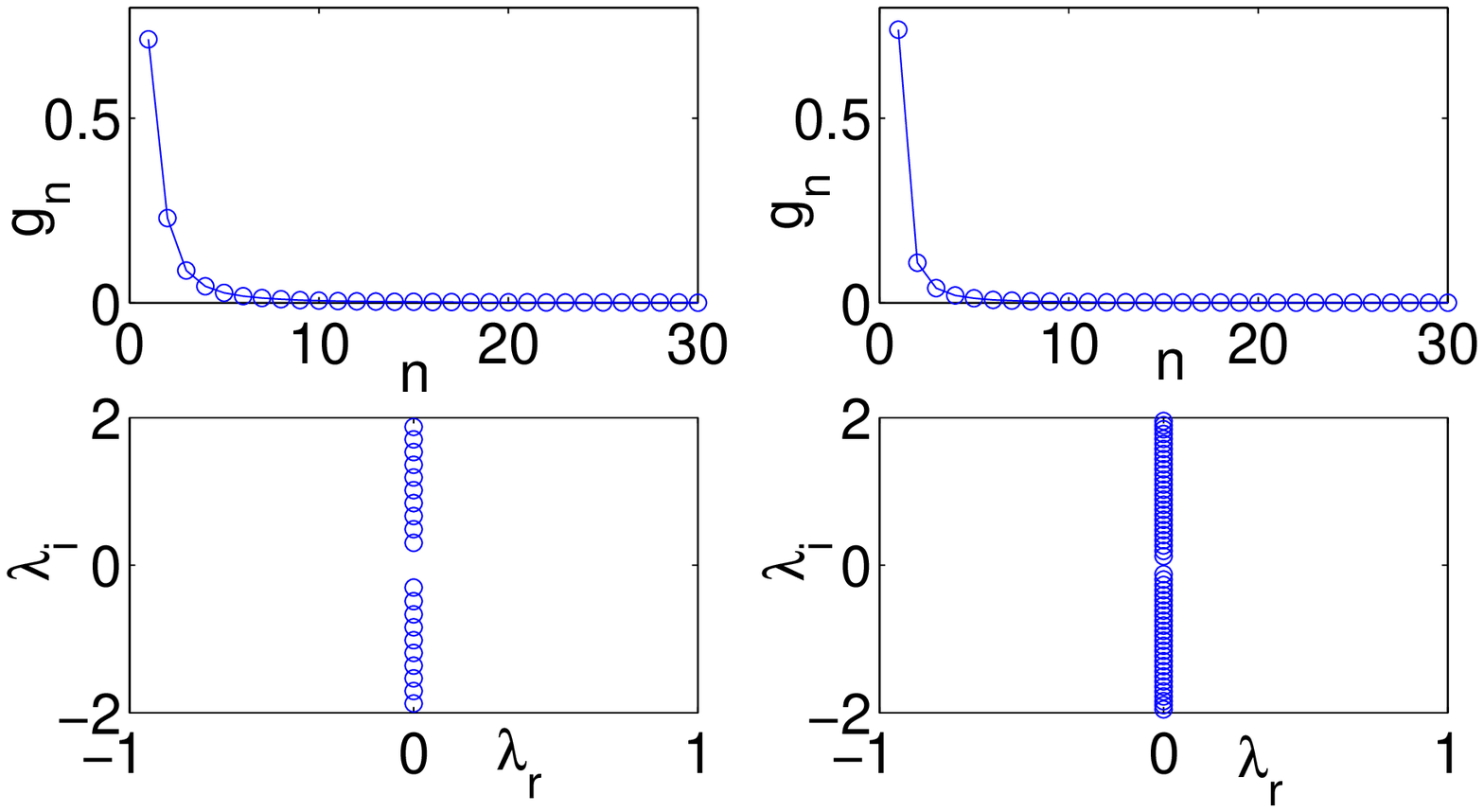}
\hfill 
\hskip .8cm
\put(0,160){b)}
\centering
 \epsfxsize=11.0cm
\epsffile{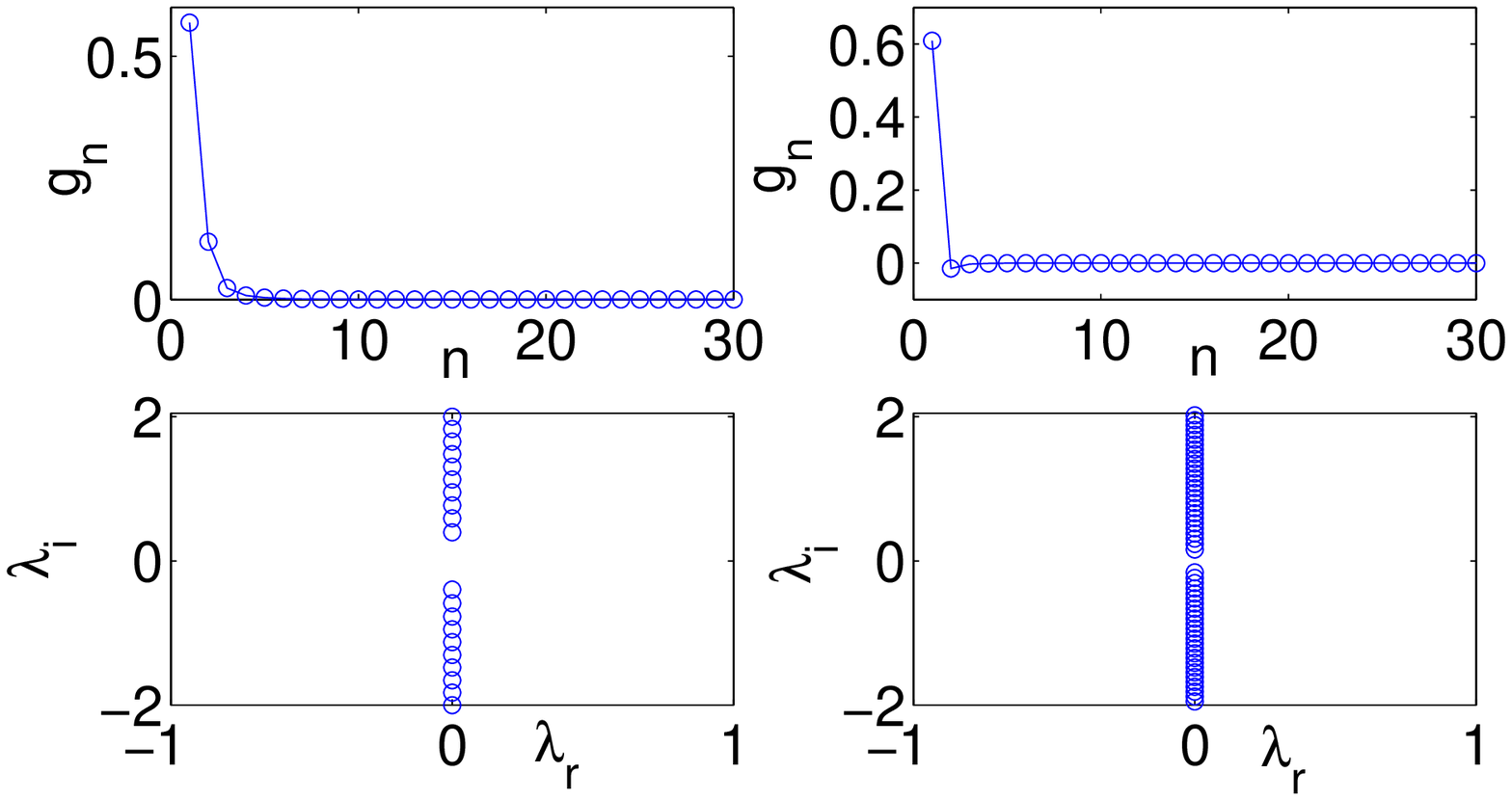}
\caption{Same as Figure \ref{fig1} but for the $3$-dimensional  case for a)
the one $SU(2)$ skyrmion  and b) for the topologically trivial $SU(3)$ 
skyrmion-antiskyrmion configuration. In each case the spatial
profile of the skyrmion and the eigenvalues of its linearization
are shown for $h=0.4$ (left panels) and $h=1$ (right panels).}
\label{fig4}
\end{figure}

The numerical existence and stability computations have been repeated 
similarly to the  $2$-dimensional case. 
The principal finding in this setting, as well, is that the
skyrmion structures are found to be linearly stable, both
in the $SU(2)$ and in the $SU(3)$ cases. In particular, 
examples of the profiles of the obtained structures and
their linear stability are illustrated in Figure \ref{fig4}
for values of $h=0.4$ (left panels) and $h=1$ (right panels)
for the two different parameter sets [top and bottom]. 
Notice the absence of any real eigenvalues showcasing the
stability of the waves.

\begin{figure}[tbp]
\begin{center}
\epsfxsize=7.0cm
\epsffile{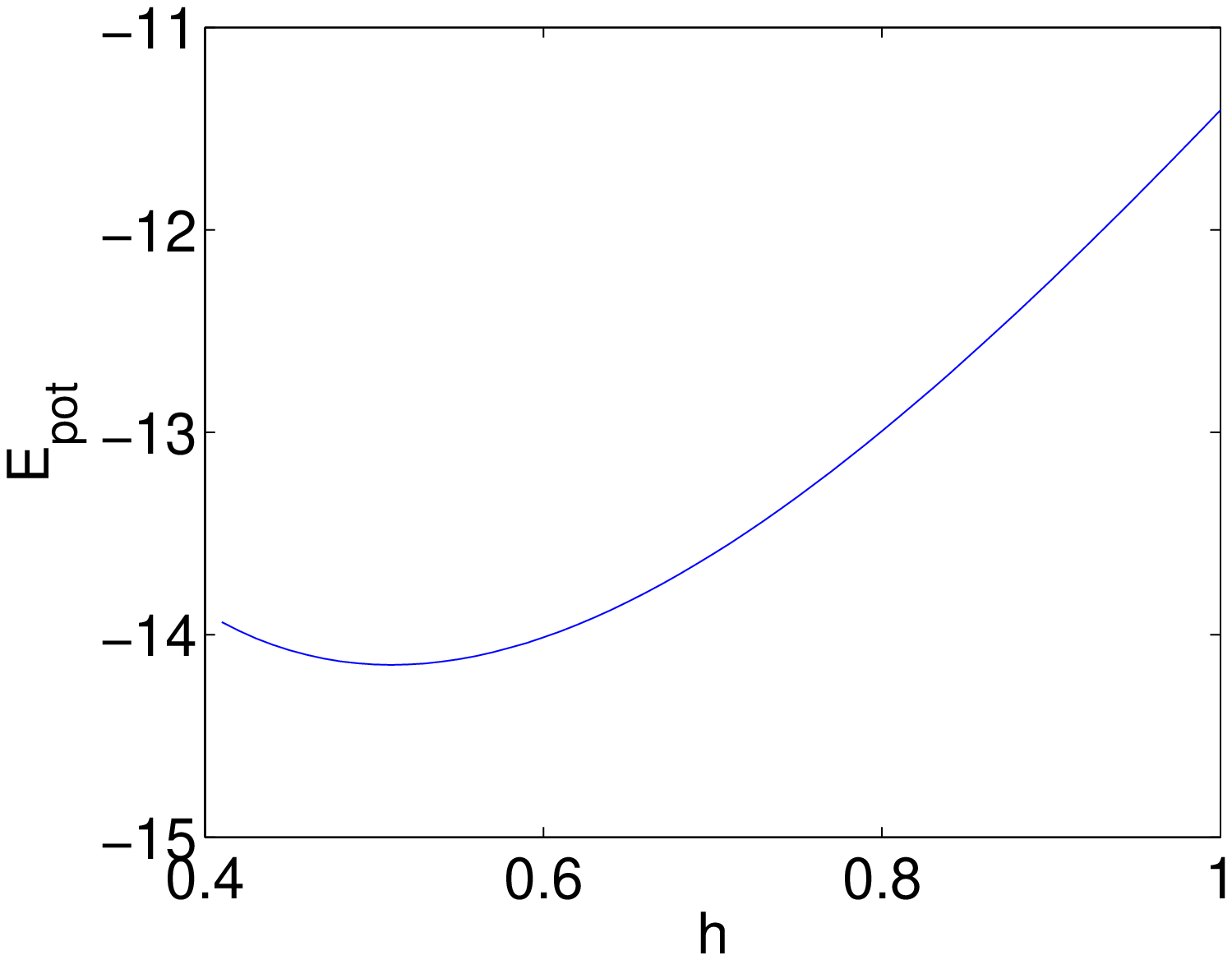}
\epsfxsize=7.0cm
\epsffile{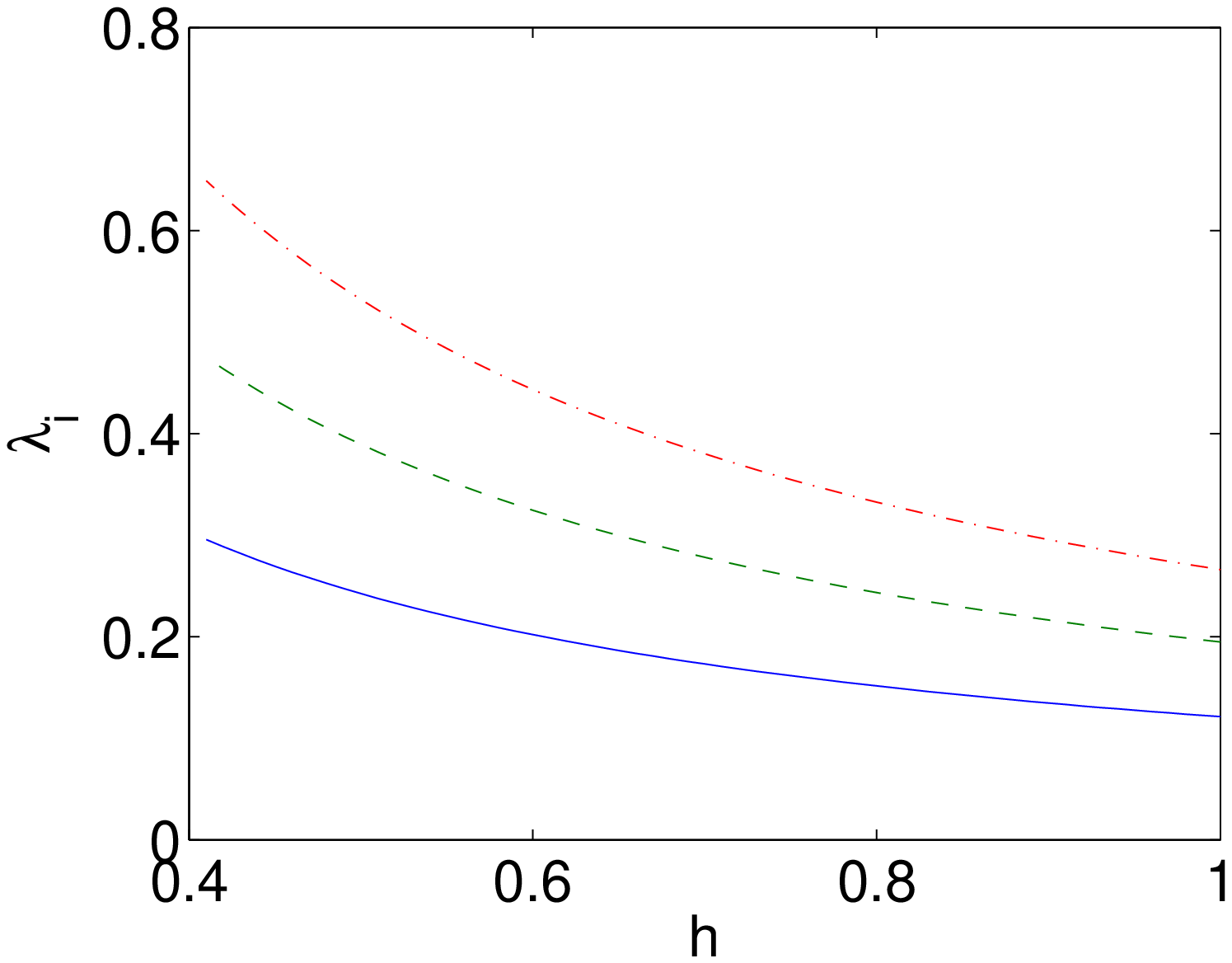}
\epsfxsize=7.0cm
\epsffile{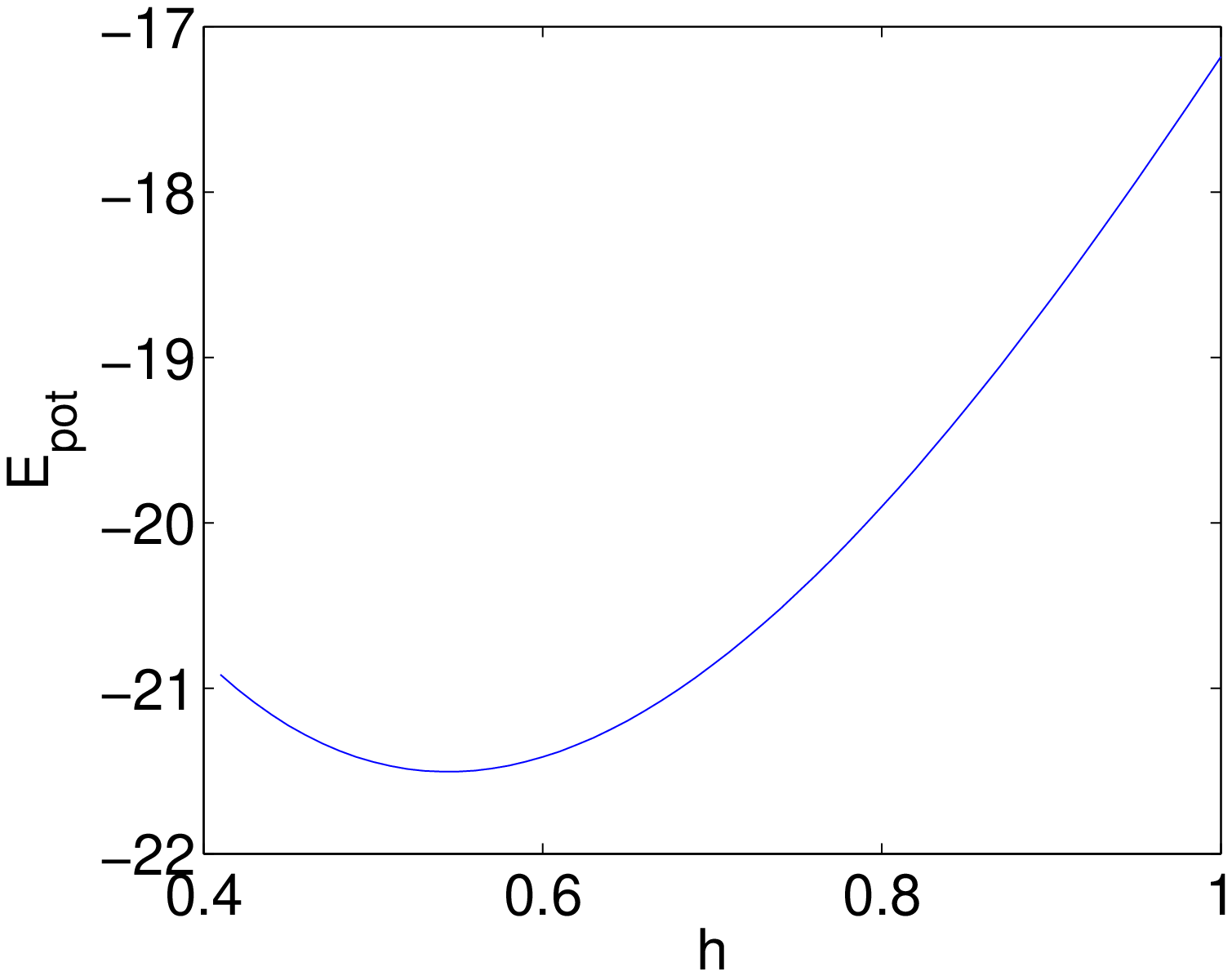}
\epsfxsize=7.0cm
\epsffile{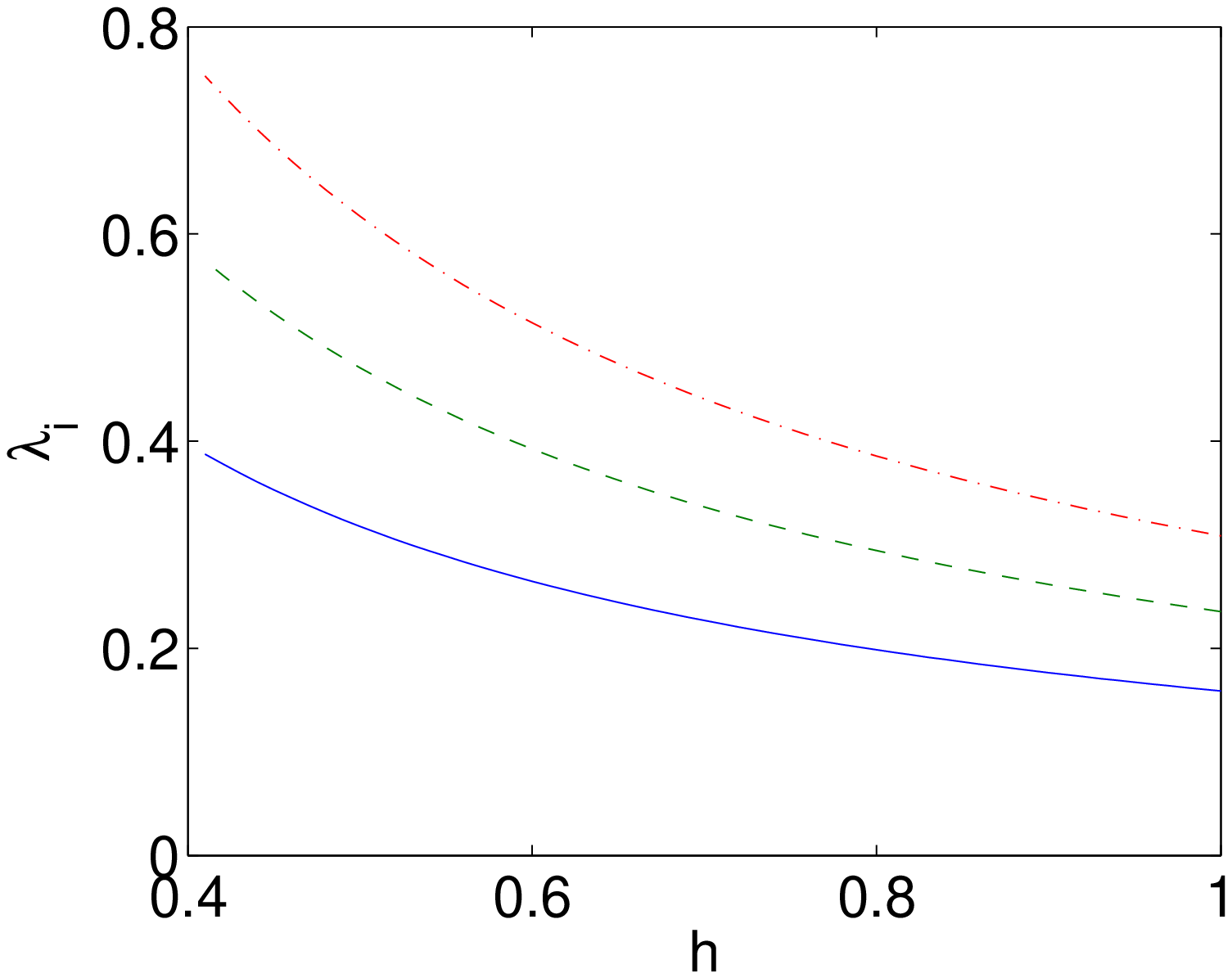}
\caption{Same as the panels of Figure \ref{fig2} but for the $3$-dimensional case
when ${\cal A}_{N}={\cal I}={\cal N}=1$ (top row) and ${\cal A}_N=4/3$ and
${\cal I}={\cal N}=4$ (bottom row). Notice that the right panels
show the three lowest eigenvalues of the linear stability in each case.}
\label{fig5}
\end{center}
\end{figure}

Using numerical continuation, with the spacing $h$ as the relevant
parameter, reveals that the structures
remain stable as $h$ is modified. The monotonically decreasing
trends of the relevant principal imaginary eigenvalues with
$h$ are shown in Figure \ref{fig5}, together with the potential
energy dependence on the spacing. The latter quantity has a non-trivial
non-monotonic dependence with a minimum around a given spacing
(which is dependent on the parameters $A_N$,  ${\cal N}$ and ${\cal I}$,
but is both cases occurring near $h=0.5$).

\begin{figure}[tbp]
\begin{center}
\epsfxsize=7.0cm
\epsffile{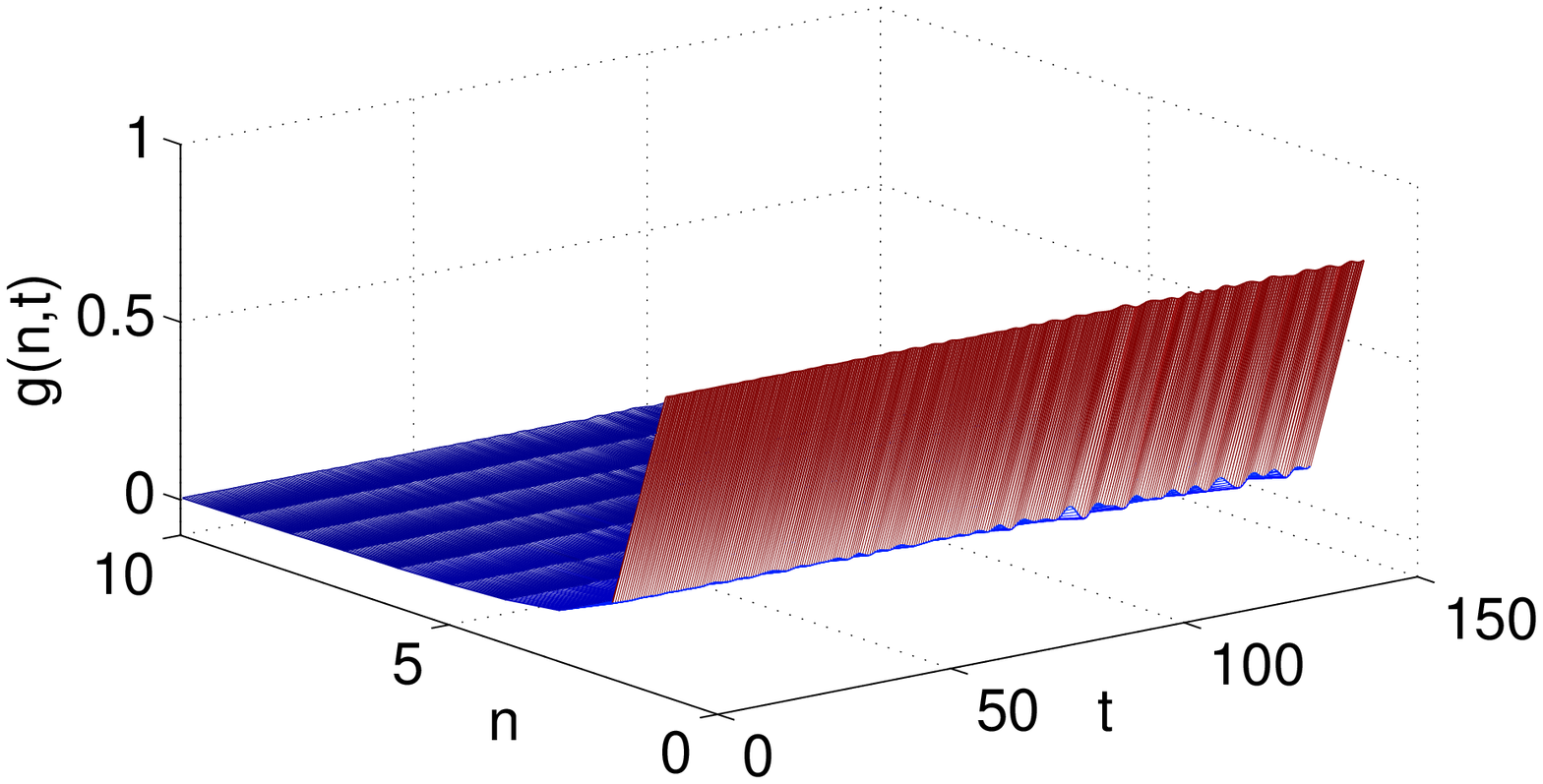}
\epsfxsize=7.0cm
\epsffile{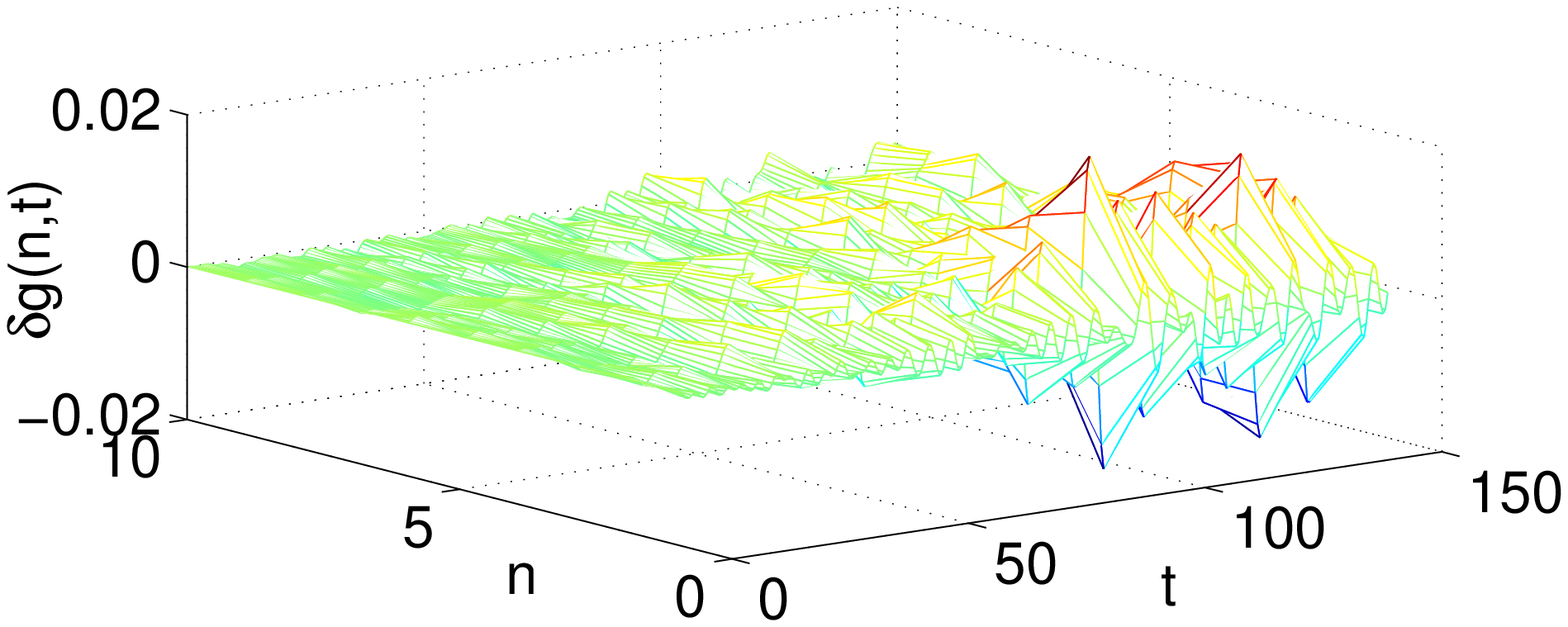}
\epsfxsize=7.0cm
\epsffile{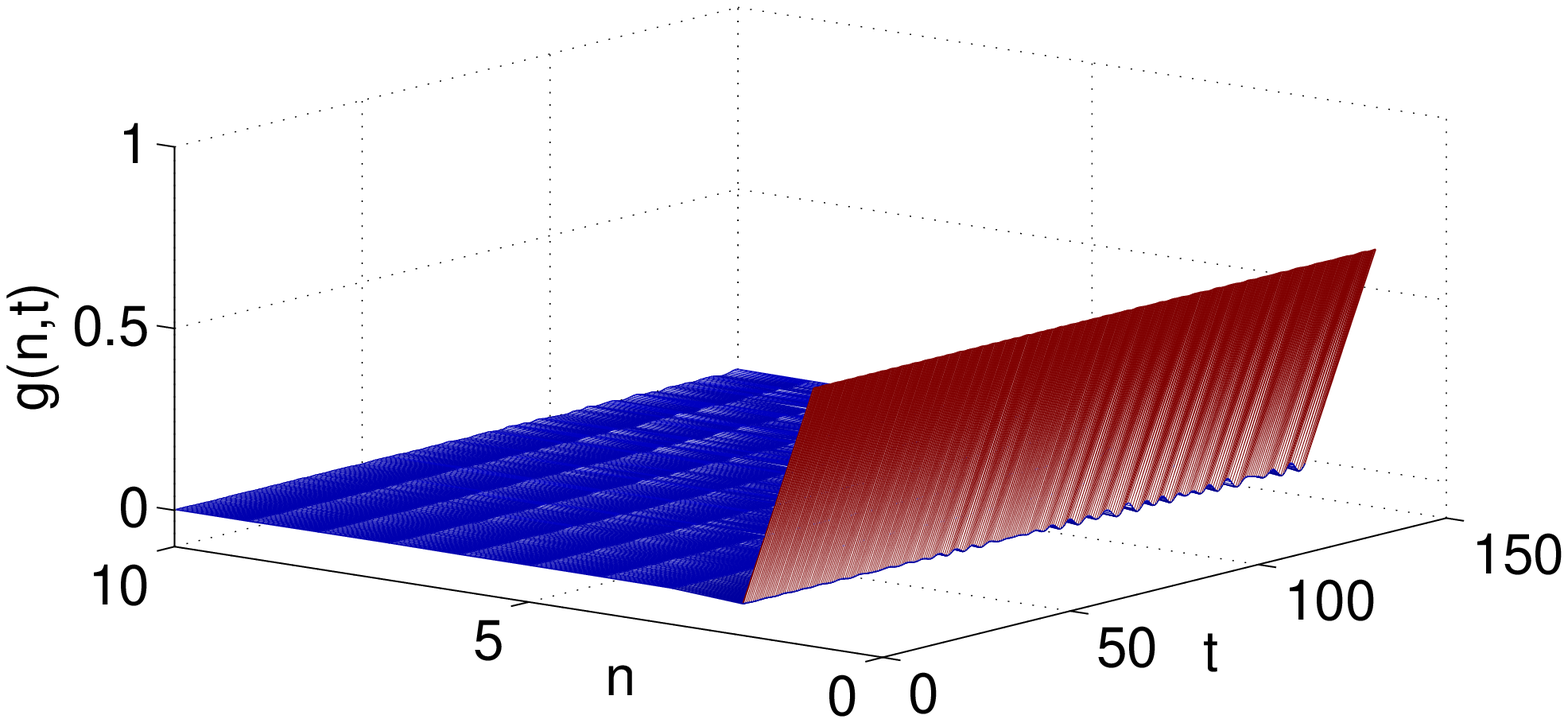}
\epsfxsize=7.0cm
\epsffile{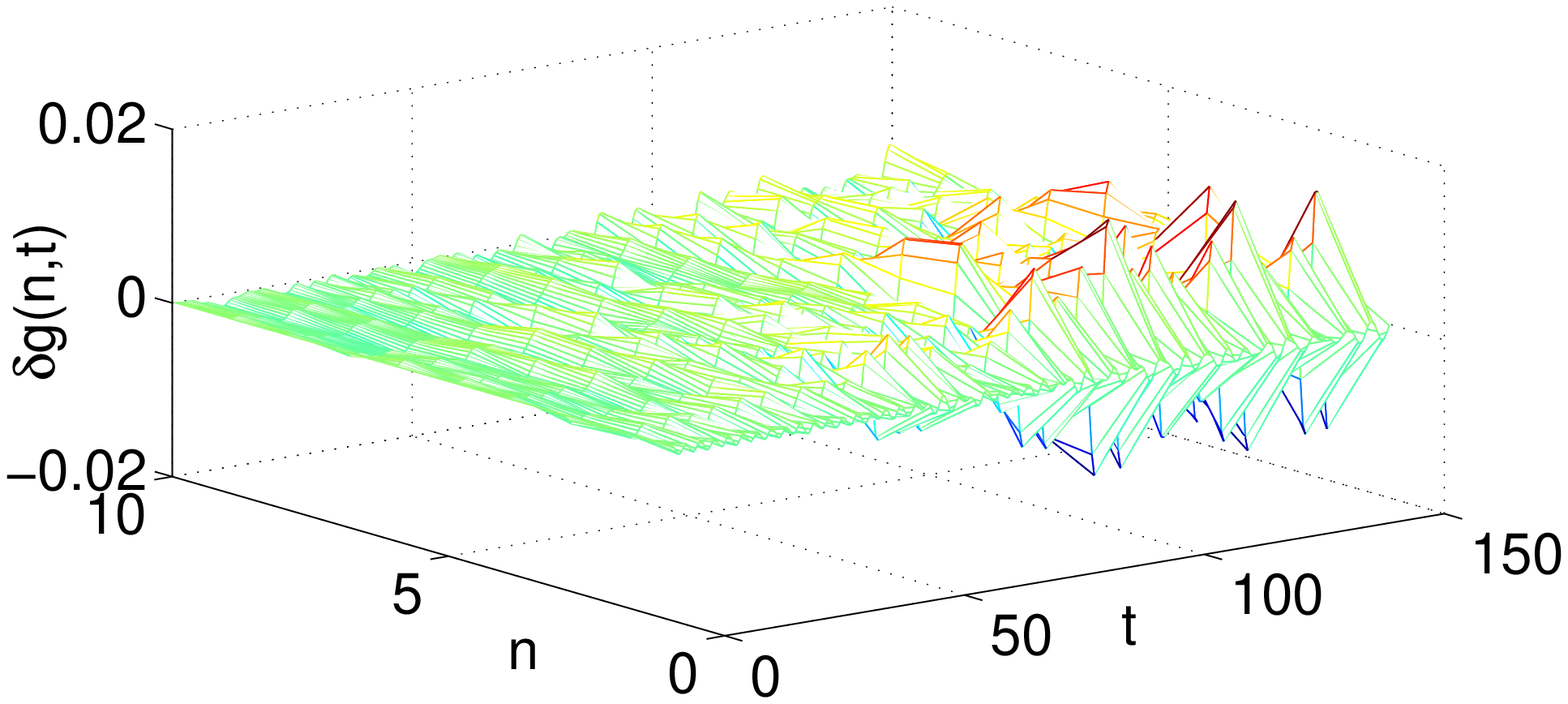}
\caption{In a format similar to that of Figure \ref{fig3}, the figure
presents direct numerical simulations of the cases with 
${\cal A}_{N}={\cal I}={\cal N}=1$ (top row) and ${\cal A}_N=4/3$ and
${\cal I}={\cal N}=4$ (bottom row). The left panel shows in both
cases the evolution of a skyrmion solution with $h=1$ in space and
time, while the right shows the deviation $\delta g(n,t)=g(n,t)-g(n,0)$
from the initial profile.}
\label{fig6}
\end{center}
\end{figure}

Finally, the numerical bifurcation analysis results were tested
against the direct numerical integration of equations (\ref{stuff3d}). These simulations are
once again particularly relevant in this context
as the dynamical evolution equation contain terms (such as
the second term in the left hand side of  (\ref{stuff3d})), which are not accounted for at the linear
stability level [arising at O$(\epsilon^2)$]; hence, it is
important to check whether linearly stable solutions may be
destabilized by such higher order effects. Our results for
these simulations are summarized in Figure \ref{fig6}, where
integration results are shown for times up to $t=150$, with
a strong perturbation to the exact solution (for $h=1$) 
of magnitude $10^{-3}$ being imposed as the initial condition. However,
the relevant perturbation remains bounded for the duration of
the simulation and even for times up to twice as large
as the ones shown here. These results clearly
indicate the robustness of the obtained solutions.

\section{Conclusions}

In this paper a novel discrete version of the Skyrme model in 
$2+1$ and $3+1$ dimensions is presented. It has been shown that both models 
admit discrete skyrmion solutions, similar to the continuum ones, which are 
well-behaved and remarkably stable, in the sense that they cannot be 
destroyed by small (or even not so small) perturbations. The discretization 
scheme is based on using polar coordinates and thus, the corresponding 
skyrmions are radially symmetric; and therefore, it is not possible to 
investigate their fully 2d or 3d dynamics. The advantage of the version 
described in this paper, on the other hand, is that the lattice spacing can 
be relatively large, without compromising the stability of the solitons.

It should be instructive to produce  discretizations of the 
full $2$- and $3$-dimensional  systems, that are not restricted to 
radially-symmetric configurations, and  to investigate
the construction of the 
corresponding skyrmions.  That way, the full dynamics of the discrete 
skyrmions can be considered, and directly tested  against  numerical 
simulations of the continuum ones \cite{Za1,Za2,BS}. The details of this
program constitute an interesting direction for future study.

{\bf Acknowledgements} 

TI thanks University of T\"ubinger for a guest Professorship position. PGK gratefully acknowledges support from grants NSF-DMS-0505663,
NSF-DMS-0619492 and NSF-CAREER.


\begin{thebibliography}{99}
\bibliographystyle{plain}
\bibitem{bog} E.B. Bogomol'nyi, Sov. J. Nucl. Phys. {\bf 24}, 449 (1976).
\bibitem{AS}
M.J. Ablowitz and H. Segur, Solitons and the Inverse Scattering Transform,
SIAM (1981).

\bibitem{L}
R. A. Leese, Phys. Rev. D 40, 2004 (1989).
\bibitem{SW}
J. M. Speight and R. S. Ward, Nonlinearity 7, 475 (1994).
\bibitem{I}
T. Ioannidou,  Nonlinearity 10,  1357 (1997).
\bibitem{ahs1} M.J. Ablowitz, B.M. Herbst and C.M. Schober,
J. Phys. A {\bf 34}, 10671 (2001).
\bibitem{fl1} S. Flach and C.R. Willis,
{\bf 295}, 181 (1998).
S. Flach and A. Gorbach, preprint (2007).
\bibitem{SM}
N. Manton and P. S. Sutcliffe, {\it Topological Solitons} (Cambridge University Press, 2004).
\bibitem{Za}
W. J. Zakrzewski, {\it Low Dimensional Sigma Models} (Adam Hilger, Bristol 1989).
\bibitem{Tom}
T. Weidig, Nonlinearity 12,  1489 (1999).
\bibitem{IKZ}
T. Ioannidou, V. B. Kopeliovich and W. J. Zakrzewski, J. Exp. Theor. Phys.
 95,  572 (2002).
\bibitem{cret} Yu.S. Kivshar, D.E. Pelinovsky,
T. Cretegny and M. Peyrard, Phys. Rev. Lett. 80, 5032 (1998).
\bibitem{pgk} P.G. Kevrekidis and C.K.R.T. Jones,
Phys. Rev. E 61, 3114 (2000).
\bibitem{HMS}
C. Houghton, N. Manton and P. Sutcliffe, Nucl. Phys. B  510, 587 (1998).
\bibitem{Ward}
R. S. Ward, Let. Math. Phys.  35,  385 (1995).
\bibitem{IPZ}
T. Ioannidou, B. Piette and W. J. Zakrzewski, J. Math. Phys.  40, 6353 (1999); J. Math. Phys.  40,  6223 (1999).
\bibitem{Za1}
M. Peyrard, B. Piette and W. J. Zakrzewski, Nonlinearity 5, 563 \&  585 (1992).
\bibitem{Za2}
B. M. A. G. Piette, B. J. Schroers and  W. J. Zakrzewski, Nucl. Phys. B 439, 205 (1995).
\bibitem{BS}
R. A. Battye and P. M. Sutcliffe, Phys. Let. B  391,  150 (1997). 
\end{thebibliography}
\end{document}